\documentclass[twoside,11pt]{report}
\usepackage{graphicx}
\usepackage{amssymb, amsmath,amsxtra}
\usepackage{epsfig,subfigure}

\usepackage{latexsym}
\usepackage{bm}
\usepackage{wrapfig}
\usepackage{cite}

\begin{document}

\begin{center} {\Large \bf Exact and approximate symmetries for light \\
  propagation equations with higher order nonlinearity}\\
[2ex] {\large Martin E. Garcia$^1$, Vladimir F. Kovalev$^2$, Larisa L.
Tatarinova$^{1,3}$}\\[1ex] {\it 1)Theoretical Physics, University of Kassel,
Heinrich-Plett-Str.\ 40, 34132 Kassel, Germany\\ 2) Institute for mathematical
modelling RAS, Miusskaya Pl., 4-A, 125047 Moscow, Russia  \\ 3) Theoretical Physics,
University of Fribourg, Chemin du Muse\'{e} 3, 1700 Fribourg, Switzerland.}
\end{center}

\noindent {\bf Abstract}\\ For the first time exact analytical solutions to the eikonal
equations in (1+1) dimensions with a refractive index being a saturated function of
intensity are constructed. It is demonstrated that the solutions exhibit collapse; an
explicit analytical expression for the self-focusing position, where the intensity
tends to infinity, is found. Based on an approximated Lie symmetry group, solutions to
the eikonal equations with arbitrary nonlinear refractive index are constructed.
Comparison between exact and approximate solutions is presented. Approximate solutions
to the nonlinear Schr\"odinger equation in (1+2) dimensions with arbitrary refractive
index and initial intensity distribution are obtained. A particular case of refractive
index consisting of Kerr refraction and multiphoton ionization is considered. It is
demonstrated that the beam collapse can take place not only at the beam axis but also
in an off-axis ring region around it. An analytical condition distinguishing these two
cases is obtained and explicit formula for the self-focusing position is presented.
\\[2ex] {\bf Keywords:} Self-focusing, nonlinear Schr\"odinger equation, eikonal
equation, symmetry group, Lie-B\"acklund symmetry.

\section{Introduction} \label{intr} 
The Lie symmetry
analysis of differential equations finds a great number of applications in mathematical
modeling of physical problems nowadays (see, e.g. Refs.
\cite{Ibragimov,K8,K9}). ).
Nonlinear optics certainly occupies a
particular place among these.

In 1960-ies it became evident that for an adequate mathematical description of the
process of highly intense light propagation the refractive index has to depend on the
intensity of applied electric field $n=n(I)$. For moderate intensities achievable at
that time it was sufficient to use the so-called Kerr form of the refractive index
$n=n_0+n_2I$ with $n_0$ being of the order of unity and $n_2$ varying from $10^{-19}$
to $10^{-16}$ in $({\rm W/cm^2})^{-1}$ depending on the material. The basic
mathematical model for the propagation of intense monochromatic light that is
successfully applied for a long time (see, e.g. classical monographs
\cite[Chapt.17]{Shen-bk} and \cite{Akhmanov}, or \cite{Boyd,Diels,CPhysRep,BergeRPP}
for recent achievements) is the nonlinear Schr\"odinger equation (NLSE) or its
approximation in the limit of geometrical optics, the eikonal equation. \par

For the first time, exact analytical solutions to the eikonal equations with Kerr-type
refractive index in (1+1) and (1+2) dimensions were constructed by Akhmanov \textit{et
al.} in Ref.~\cite{AkhmanovJETP,Akhmanov}. The authors demonstrated that in both cases
the solutions exhibit singularities at certain points and found explicit analytical
expressions for them. Later in Ref.~\cite{KovPust,Kovalev1+1} it was demonstrated that
these solutions can be derived in a regular manner using the Lie-B\"acklund symmetry
group admitted by the eikonal equation with Kerr refractive index. \par

Lie symmetry group analysis of NLSE has been performed by many authors (see,
e.g.~\cite[Chap.16]{Ibragimov}). In particular, L. Gagnon and P. Winternitz in Ref.
\cite{Gagnon,Gagnon1,Gagnon2} found exact solutions in (1+2) dimensions, however, these
solution did not correspond to localized (symmetric) intensity distributions typical
for usual experimental conditions. A set of two coupled NLSE was analyzed by means of
Lie group technique and the general Lie group of point symmetries, its Lie algebra, and
a group of adjoint representations that corresponds to the Lie algebra were identified
in Ref \cite{Bolgary}.

For the initial conditions actual for typical experiments, analytical solutions to the
eikonal equation in (1+1) dimensions based on the symmetry group approach were obtained
in Ref. \cite{Kovalev1+1}. In this paper approximate solutions for various initial
intensity distributions and Kerr-type media were constructed. Later, using the similar
group analysis technique approximate solutions in (1+2) dimensions for arbitrary
initial intensity profile and the same form of the refractive index were found in Refs.
\cite{Kovalev1+2,KovalevPRA}. Based of the obtained solutions, the authors proceeded to
investigate a global behavior of the solutions and to get explicit analytical
expressions for the nonlinear self-focusing position and the value of critical power
required for beam collapse.

On the other hand, modern experimental facilities allow one to achieve very intense
laser beams leading to highly nonlinear media response. In such situations the Kerr
approximation to the refractive index ceases to be sufficient and higher order terms
with respect to power of the light intensity must be taken into account.

In the present paper we study the problem of light propagation in media with highly
nonlinear response. Based on Lie symmetry group analysis, we constructed an exact
solution to the eikonal equation in (1+1) dimensions for a special higher-order form of
the refractive index, and also approximate analytical solutions to the problem in both
(1+1) and (1+2) dimensions with the refractive index being an arbitrary function of the
intensity.

The paper is organized as follows. First, after describing the model equations, we
consider the problem of light propagation in (1+1) dimensions under an approximation of
geometrical optics. Based of the formalism of Ref. \cite{KovPust} we construct the
Lie-B\"acklund symmetry group admitted by the eikonal equation and consider such a
superposition of the symmetry operators that yields a localized initial light intensity
distribution. The use of this combination of operators gives us an exact solution of
the eikonal equation with the refractive index being a saturated function of the
intensity of the applied electric field. The solution exhibits a singularity: the
on-axial intensity asymptotically tends to infinity at a certain propagation distance.

In the next section we construct an approximate analytical solution to the eikonal
equation in (1+1) dimensions with arbitrary nonlinear media response. This solution is
obtained on the basis of the approximate Lie-B\"acklund symmetry group. In order to
test the applicability of used approximation the solutions obtained on the basis of
exact and approximate Lie symmetry groups under the same initial conditions are
compared.

The last section is devoted to the construction of approximate analytical solutions of
the Schr\"odinger equation in (1+2) dimensions when both the refractive index and the
initial intensity distributions are arbitrary. The final result is a system of
algebraic equations which has to be resolved for every particular initial condition and
nonlinear media response. The most typical situation in modern experiments (see e.g.
\cite{CPhysRep,BergeRPP}) is the propagation of the Gaussian beam in an ionizing media,
with the refractive index being a polynomial function of the light intensity.
Therefore, we consider this problem as a particular application of the obtained
results. Influence of the higher order nonlinear term in the refractive index on the
beam collapse is considered and result is compared with the previous one obtained in
Refs. \cite{Kovalev1+2,KovalevPRA}.

\section{Model equations} Let us start from the NLSE: \begin{eqnarray} i {\cal E}_z
+{1\over 2k_0}\nabla_{\bot}^2 {\cal E} +k_0n(\vert {\cal E} \vert^2) {\cal E} = 0
\,.\label{NLSE} \end{eqnarray} Here ${\cal E}$ is the slowly-varying envelope of the
electric field, $z$ is the propagation length, $k_0$ is the wave number
$k_0=n_0\omega_0/c$, $\omega_0$ is the carrier frequency of the laser irradiation, $c$
is the velocity of light and  $n = n(\vert{\cal E}\vert^2)$ is a nonlinear refractive
index in a general form (see e.g. \cite{Boyd,CPhysRep,BergeRPP}). Due to high
intensities of light available in modern experiments the refractive index becomes
highly nonlinear $n = n(\vert{\cal E}\vert^2)$. The Laplace operator $\nabla_{\bot}^2$
is usually responsible for light diffraction. Explicitly,
$\nabla_{\bot}^2=\partial_{xx}$ or $\partial_{xx}+\partial_{yy}$ for (1+1) or (1+2)
dimensional cases correspondingly.

Let us now represent electric field ${\cal E}$ in the eikonal form: ${\cal
E}=\sqrt{I}\exp(ik_0Q)$. Then, starting from Eq.~(\ref{NLSE}), after some algebraic
manipulations we obtain \begin{eqnarray} && Q_z=-{1\over 2}(Q_x)^2+ n(I) +
\frac{1}{2k_0^2} \left({x^{1-\nu}
\over\sqrt{I}}\partial_{x}(x^{\nu-1}\partial_{x}\sqrt{I}) \right), \label{BE0a}\\ &&
I_z= - \partial_{x}( I Q_x)-(\nu-1){I
 Q_x\over {x}}, \label{BE0b}
\end{eqnarray} where $\nu=1$ and $\nu=2$ correspond to the (1+1) and (1+2) dimensional
cases, and $x$ denotes the transverse spatial variable.

Let us differentiate the first equation with respect to $x$ and introduce a new
variable $v \equiv Q_x$. For the sake of convenience, we introduce dimensionless
variables $\tilde{I}\equiv I/I_0$, $\tilde{x}\equiv x/w_{\rm in}$, $\tilde{z}\equiv
z/w_{\rm in}$, where $I_0$ is an initial peak intensity of the light beam and $w_{\rm
in}$ is an initial beam radius.  In what follows the dimensionless parameters shall
always be used, omitting the tilde for simplicity. Moreover, let us introduce new
dimensionless variables $\alpha=n_2 I_0$, $\theta= (2k_0^2 w_{\rm in}^2)^{-1}$ in
diffractive case. Thus, finally we get the following equations: \begin{eqnarray}
 && v_z + v v_{x} - \alpha\varphi I_{x}-\theta\partial_{x}
 \left({{x}^{1-\nu} \over\sqrt{I}}\partial_{x}\left(
 x^{\nu-1}\partial_{x}\sqrt{I}\right) \right)=0,
 \label{BE1a}\\
 && I_z + v I_{x}+I v_{x} +(\nu-1){v I\over
 x}=0, \quad \varphi=\partial_I n \, . \label{BE1b}
\end{eqnarray} Evidently, Eqs. (\ref{BE1a}), (\ref{BE1b}) must be supplemented with a
boundary conditions. In case of collimated beam these read \begin{eqnarray}
v(0,x)=0,\hspace{2cm} I(0,x)=I_0(x), \end{eqnarray} In several cases the term with
higher order derivatives can also be neglected Refs. \cite{TatGar2} and equations
(\ref{BE1a}), (\ref{BE1b}) acquire a rather simple form: \begin{equation}
 \begin{aligned}
 & v_z + v v_x-\alpha \varphi I_x=0 \,, \quad  I_z + v I_x + I v_x + (\nu-1){I v\over
 x}=0.
 \end{aligned}
 \label{BE1}
\end{equation} Further simplification of these equations can be performed in (1+1)
dimensions if one notices that in this case the system (\ref{BE1}) is linear with
respect to the first order derivatives. Therefore, it is convenient to use the
hodograph transformation in order to transform it into a linear system of partial
differential equations. In doing so, in (1+1) dimensions one obtains \begin{equation}
\begin{aligned}
 & \alpha \tau_v-{I\over\varphi(I)} \chi_I = 0, \qquad  \chi_v + \tau_I = 0 \,.
 \end{aligned}
 \label{hod1}
\end{equation} In (1+$\nu$) dimensions the Eqs.~(\ref{BE1}) read \begin{equation}
\begin{aligned}
 & \alpha \tau_v -{I\over\varphi(I)}\chi_I=0 \,,  \\
 & \chi_v + \tau_I + \frac{(\nu-1) v}{ \chi I+ \tau v} \left[
 \chi_v \left( I \tau_I - \tau \right) + \tau \tau_I - \frac{\tau^2}{I}
 - \chi_I^2 \frac{I^2}{\varphi(I)} \right]=0\,. \end{aligned}
 \label{hod1nu}
\end{equation} The boundary conditions are transformed as follows: for $v=0$
\begin{equation} \tau=0\,, \qquad \chi=H(I) \,,
 \label{hodBC}
\end{equation} where $H(I)$ is a function inverse to a smooth initial intensity
distribution $I_0=I_0(x)$. Evidently, for example, in case of a Gaussian beam we have
$\chi=\sqrt{\ln(1/I)}$.

\section{An exact solution to the eikonal equations in (1+1) dimensions} Let us now
construct a new exact analytical solution to Eqs.~(\ref{hod1}). In the present paper,
for the first time, we consider a special form of the nonlinear media response with the
refractive index being a saturated function of intensity. Let us rewrite
Eqs.~(\ref{hod1}) as follows: \begin{eqnarray} \tau_v - \psi(I)\chi_I = 0,\hspace{1cm}
\chi_v + \tau_I = 0,\label{BVP} \end{eqnarray} where $\psi = I/(\alpha \varphi)$. As a
boundary condition, we take a collimated continuous wave beam with a localized
symmetric intensity distribution at the entry plane of a nonlinear media
\begin{eqnarray} \tau(I,0) = 0,\hspace{1cm}\chi(I,0) = \chi_0(I).\label{BC}
\end{eqnarray}

\begin{figure}[!h]
 \centering
 \includegraphics[angle=0,width=9cm]{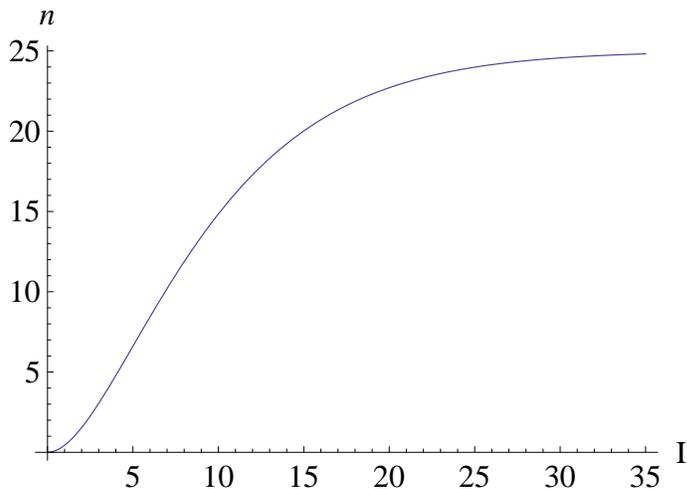}
 \caption{\label{n(I)} Nonlinear refractive index corresponding to
 $\psi={\rm e}^{bI}/\alpha$ as a function of the normalized intensity
 $I$.}
\end{figure}

Our goal is now to construct an exact analytical solution to the system of equations
(\ref{BVP}) with a nonlinear function $\psi(I)$ that corresponds to a saturating
dependence of the refractive index on the intensity. For this goal, $\psi$ is taken in
the form ${\rm e}^{bI/I_0}/n_2I_0$. The refractive index corresponding to this choice
of $\psi$ is presented in Fig.~\ref{n(I)}. 

Let us first sketch the broad outlines of our solution: first, we construct a Lie
symmetry group admitted by Eqs.~(\ref{BVP}). Second, following a formal scheme reported
in Refs.~\cite{ShirKovPhysRep}, the obtained group shall be restricted to the surface
of boundary conditions: $v = 0$, $\tau = 0$. All derivatives of $\tau$ with respect to
$I$ have to vanish too. Third, based on the requirement of vanishing of canonical
coordinates of the group generators on the boundary, we shall construct such a linear
superposition of them, which provides a localized beam intensity distribution. Finally,
the integration of the constructed superposition shall yield a desired solution to
Eqs.~(\ref{BVP}).

\subsection{Recursion operators and Lie-B\"acklund symmetries of the second order} We
start from the search for the Lie-B\"acklund symmetry group admissible by
Eqs.~(\ref{BVP}). It is generated by the canonical infinitesimal operators
\cite{Ibragimov} \begin{equation}
 X = f^s \partial_{\tau} + g^s \partial_{\chi}\,, \label{X}
\end{equation} with coordinates $f^s$ and $g^s$. For the Lie-B\"acklund symmetry of
arbitrary order $s>1$ the coordinates $f^s$ and $g^s$ depend on $v$, $I$, $\tau$,
$\chi$ and corresponding derivatives of $\tau$ and $\chi$ up to the $s$-th order with
respect to $I$ \[
 f^s = f^s(v, I, \tau, \chi,..., \tau_I^s,\chi_I^s),\quad
 g^s = g^s(v, I, \tau, \chi,..., \tau_I^s,\chi_I^s)\,.
\] Here, the index $s$ stands for the order of the derivatives: $\tau_I^s\equiv
\partial^s \tau/\partial I^s$, etc. The coordinates $f^s$ and $g^s$ are found from the
determining equations that in the case of Eqs.~(\ref{BVP}) read \cite{KovPust}:
\begin{equation}
 D_v(f^s) - \psi D_I(g^s) = 0, \qquad D_v(g^s) + D_I(f^s) = 0 \,,
 \label{DE}
\end{equation} where $D_I$ and $D_v$ are operators of the total differentiation with
respect to $I$ and $v$: \begin{equation} \begin{aligned} &
 D_I = \partial_I +
 \sum_{s=0}^{\infty} \left( \tau_{I}^{s+1}\partial_{\tau_I^s} +
 \chi_{I}^{s+1}\partial_{\chi_I^s} \right), \\ &
 D_v = \partial_v + \sum_{s=0}^{\infty} \left[ \left( \psi \chi_I \right)_{I^s}
 \partial_{\tau_I^s}
 - \tau_{I}^{s+1} \partial_{\chi_I^s} \right] \,.
 \end{aligned}
 \label{Dn}
\end{equation} In order to solve the Eqs.~(\ref{DE}), it appears more convenient to use
a recursion operator \cite{KovPust}. The latter is defined as $2\times 2$ matrix
operator transforming any linear solution of the determining equation (\ref{DE}) of the
order $s$ to the solution of these equations of higher order $(s+1)$ \begin{eqnarray}
L\left( \begin{array}{c} f^s \\ g^s \end{array} \right) = \left( \begin{array}{c}
f^{s+1} \\ g^{s+1} \end{array} \right), \hspace{1cm}L = \left( \begin{array}{cc} L^{11}
& L^{12} \\ L^{21} & L^{22} \end{array} \right),\label{L} \end{eqnarray}

Substitution of Eq.~(\ref{L}) into the determining equation (\ref{DE}) yields the
following system of equations for the elements of $L$ \begin{equation} \begin{aligned}
&
 (D_vL^{11} - \psi D_I L^{21})f^s + (D_vL^{12} - \psi D_I L^{22})g^s = 0\,,
 \\ &
 (D_IL^{11} + D_vL^{21})f^s + (D_I L^{12} + D_v L^{22})g^s = 0,
 \end{aligned}
 \label{DEL}
\end{equation} which should be valid for any solutions $f^s$ and $g^s$.

Explicit formulae for the components $L^{ij}$ $(i,j=1...3)$ of recursion operators
(\ref{L}) are given in Ref.~\cite{KovPust} \begin{eqnarray}
 L_1 &=& \left( \begin{array}{cc} 0 & - \psi D_I/\alpha \\ D_I &
 0\end{array} \right), \quad 
 L_2 \, \, = \, \, \left( \begin{array}{cc} 2\sigma D_I-1 & -\psi(1-2\sigma_I)v D_I \\
 (1-2\sigma_I)vD_I & 2\sigma D_I \end{array} \right),\nonumber\\
 & & \qquad \qquad L_3 \, \, =\, \, \left( \begin{array}{cc} 2\sigma v D_I -
 (1-\sigma_I)v &
 -\psi q D_I - \sigma \\ qD_I + \psi_I^{-1} & 2\sigma vD_I +
 v\sigma_I \end{array} \right),\nonumber
\end{eqnarray} where $\sigma\equiv\psi/\psi_I$ and $q\equiv (1-2\sigma_I)v^2/2 + 2\int
\psi_I^{-1}dI$. The formula for the recursion operator $L_1$ is valid for arbitrary
nonlinearity function $\psi(I)$, while operators $L_2$ and $L_3$ arise for those
functions $\psi(I)$ that fulfill the condition Ref. \cite{KovPust}: \begin{eqnarray}
\left({\psi\over \psi_I}\right)_{II} = 0.\label{Cond} \end{eqnarray} It has been just
this requirement which has defined our particular choice of the function $\psi(I)$ in
initial statement of the problem. For this form of $\psi$ we have:
\begin{eqnarray}
\varphi=I {\rm e}^{-bI},\hspace{1cm}\sigma = b^{-1},\hspace{1cm} q = v^2/2 -
2\alpha{\rm e}^{-bI}/b^2.\nonumber
\end{eqnarray}
In this particular case, the
recursion operators read:
\begin{eqnarray}
 L_1 &=& \left(
  \begin{array}{cc}
  0   & -{\rm e}^{bI}D_I/\alpha \\
  D_I & 0
  \end{array} \right), 
 \quad L_2  \, = \, \left( \begin{array}{cc}
  2D_I/b-1 & -{\rm e}^{bI}v D_I/\alpha \\
  vD_I     & 2D_I/b \end{array} \right),\nonumber\\
 L_3 &=& \left( \begin{array}{cc}
  2vD_I/b-v &  \\
  \left(v^2/2 - 2\alpha {\rm e}^{-bI}/b^2\right)D_I + {\rm e}^{-bI}\alpha/  b &
  \end{array} \right. \nonumber \\
  & & \left.  \begin{array}{cc}
    &\hspace{3cm}  -{\rm e}^{bI}\left(v^2/2 - 2\alpha {\rm e}^{-bI}/b^2
    \right)D_I/\alpha-1/b \\
    & 2vD_I/b
  \end{array} \right).\nonumber
\end{eqnarray} Let us now proceed with constructing the Lie-point symmetry group
admitted by Eqs.~(\ref{BVP}) on the basis of these operators.

An evident solution of the determining equation (\ref{DE}) is \begin{equation}
 f_0^0 = \tau, \qquad g_0^0 = \chi\,.\label{f0g0}
\end{equation} The action of three recursion operators $L_i,\hspace{0.2cm} i = 1,2,3$
on the vector with coordinates given by Eq.~(\ref{f0g0}) in accordance with
Eq.~(\ref{L}) generates the symmetry group given by
\begin{equation}
\begin{aligned} &
 f_1^1 = -{\rm e}^{bI}\chi_I/\alpha\,, \quad
 f_2^1 =  \frac{2\tau_I}{b}
 - \tau - \frac{{\rm e}^{bI}v\chi_I}{\alpha}\,, \\ &
 f_3^1 = \frac{2v\tau_I}{b} - v\tau - \frac{{\rm e}^{bI}}{\alpha}
 \left( \frac{v^2}{2} - \frac{2\alpha {\rm e}^{-bI}}{b^2}\right)\chi_I
 - \frac{\chi}{b}, \\ &
 g_1^1 = \tau_I \,, \quad
 g_2^1 = v\tau_I + 2\chi_I/b \, , \\ &
 g_3^1 = \left(\frac{v^2}{2} - \frac{2\alpha
 {\rm e}^{-bI}}{b^2}\right)\tau_I + \frac{\alpha {\rm
 e}^{-bI}\tau}{b} + \frac{2v\chi_I}{b}\,.
\end{aligned} \label{point} \end{equation} Admissible for arbitrary nonlinearity
$\psi(I)$, the symmetries $f_0^0$, $g_0^0$ describe the dilatation of $\tau$ and
$\chi$, whilst $f_1^1$ and $g_1^1$ generate translations along $v$-axis.

As it was formulated in Refs.~\cite{KovPust,ShirKovPhysRep}, an invariant solution to
the boundary value problem, in particular the one given by Eqs.~(\ref{BVP}), must be
found from the constructed Lie-B\"acklund symmetries under the invariance conditions
\begin{eqnarray} f = 0,\qquad g = 0\,,\label{InvCond} \end{eqnarray} supplemented by
the original Eqs~(\ref{BVP}). In Eqs.~(\ref{InvCond}) the functions $f$ and $g$ are
arbitrary linear combinations of coordinates $f_i^s$ and $g_i^s$ of the group
generators Eqs.~(\ref{point}) and must be chosen to satisfy the boundary conditions
what in the actual case provides a localized intensity distribution.

Unfortunately, the Lie point group generators (\ref{point}) with $s=0$ and $s=1$ are
not sufficient in order to determine a linear superposition able to satisfy a smooth
localized (symmetric in $\chi$) intensity distribution at the boundary. Therefore, we
shall continue using the discussed approach with operators $L_i$ given by
Eqs.~(\ref{L}) and the vectors with coordinates of Eq.~(\ref{point}) in order to find
the Lie-B\"acklund symmetries of the higher order with $s>1$. However, since the
further calculations are quite cumbersome, it is convenient first to find the symmetry
coordinates at the boundary where they have the simplest form. Afterwards, we
completely reconstruct only those that will be included into the chosen linear
superposition.

Thus, at the boundary $\tau=0$, $v=0$ the recursion operators read \begin{eqnarray} L_1
= \left( \begin{array}{cc} 0 & -{\rm e}^{bI}D_I/\alpha \\*
      D_I & 0 \end{array} \right), \hspace{1cm}
L_2 = \left( \begin{array}{cc} 2D_I/b-1 & 0 \\
      0 & 2D_I/b \end{array} \right),\\*
L_3 = \left( \begin{array}{cc} 0 & 2/b^2 D_I-1/b \\
     -2\alpha {\rm e}^{-bI}/b^2D_I + {\rm e}^{-bI}\alpha/  b & 0
     \end{array} \right),
\end{eqnarray} Action of these operators on Eqs.~(\ref{f0g0}) gives 9 symmetry
operators $X_i$ whose coordinates $f_i^s$, $g_i^s$ are listed in the Table
\ref{fgbound}.

\begin{table}\caption{\label{fgbound}Coordinates of the symmetry operators at the
boundary.} \begin{tabular}{|c|c|c|}\hline
     & $f_i$ & $g_i$  \\ \hline
$i=0$& $0$   & $\chi$ \\ \hline $i=1$& $-{\rm e}^{bI}\chi_I/\alpha$ & $0$ \\ $i=2$& $0$
& $2\chi_I/b$ \\ $i=3$& $2\chi_I/b^2-\chi/b$ & $0$ \\ \hline $i=4$& $-2{\rm
e}^{bI}\chi_{II}/\alpha b$ & $0$ \\ $i=5$& $0$ & $-b{\rm e}^{bI}\chi_I/\alpha - {\rm
e}^{bI}\chi_{II}/\alpha$ \\ $i=6$& $0$ & $2\chi_{II}/b^2 - \chi_I/b$  \\ $i=7$& $0$ &
$4\chi_{II}/b^2$  \\ $i=8$& $-2{\rm e}^{bI}\chi_{II}/\alpha b - {\rm
e}^{bI}\chi_I/\alpha$ & $0$ \\ $i=9$& $4\chi_{II}/b^3 - 4\chi_I/b^2 + \chi/b$& $0$  \\
$i=10$& $4\chi_{II}/b^3 - 2\chi_I/b^2$ & $0$  \\ $i=11$& $0$ & $2\chi_{II}/b^2 +
\chi_I/b$ \\ $i=12$& $0$ & $-4\alpha {\rm e}^{-bI}\chi_{II}/b^4 + 4\alpha {\rm
e}^{-bI}\chi_I/b^3 -
               \alpha\chi {\rm e}^{-bI}/b^2$  \\ \hline
\end{tabular} \end{table}

Based on the operators presented in the Table \ref{fgbound}, one can construct a linear
superposition providing a localized intensity distribution. For instance, the equation
\begin{eqnarray} (2 - {\rm e}^{bI-1})\chi_{II} + b(1-{\rm e}^{bI-1})\chi_I =
0\label{Eq} \end{eqnarray} has a particular solution $\chi = \sqrt{2{\rm e}^{-bI+1} -
1}$. Resolving $I$ as a function of $\chi$, we get a convex symmetric on $x$ intensity
distribution \begin{equation}
 I_0(\chi) = \frac{1}{b}\left(1 - \ln\left( \frac{\chi^2 + 1}{2} \right)\right)\,,
 \label{PS}
\end{equation} which is represented by a red curve in Fig.~\ref{I_0}.

From the Table \ref{fgbound} one can see that Eq.~(\ref{Eq}) corresponds to the
following superposition of symmetry operators of the first and the second order
\begin{equation}
 \frac{\alpha}{{\rm e}}g_5^2 + \frac{b^2}{2}g_7^2 +
 \frac{b^2}{2}g_2^1 = 0,
\label{g} \end{equation} which, evidently, shall be supplemented by the equation:
\begin{eqnarray} {\alpha \over {\rm e}}f_5^2 + {b^2\over 2}f_7^2 + {b^2\over 2}f_2^1 =
0.\label{f} \end{eqnarray} It is easy to see that in order to find an invariant
solution satisfying Eqs.~(\ref{BVP}) and the boundary conditions of Eq.~(\ref{PS}), we
have to reconstruct a complete form of the symmetry coordinates $f_5^2$, $g_5^2$ and
$f_7^2$, $g_7^2$. Acting by the operator $L_1$ on the couple $f_1^1$, $g_1^1$, we get
\begin{equation}
 \begin{aligned} &
  f_5^2 = - {\rm e}^{bI}  \tau_{II}/\alpha \,,  \\ &
  g_5^2 = - b{\rm e}^{bI}(\chi_{II} + b\chi_I)/\alpha \,.
 \end{aligned}
\label{f5g5} \end{equation} The similar procedure applied to the operator $L_2$ and
coordinates $f_2^1$, $g_2^1$ yields: \begin{equation}
 \begin{aligned} &
  f_7^2 = (4\alpha - {\rm e}^{bI}v^2b^2)\tau_{II}/ \alpha b^2 - 4\tau_I/b -
       {\rm e}^{bI}v\chi_I/\alpha - 4{\rm e}^{bI}v\chi_{II}/\alpha b + \tau
   \, , \\ &
 g_7^2 = 4v\tau_{II}/b - v\tau_I - {\rm e}^{bI}bv^2\chi_I/\alpha +
 (4\alpha - {\rm e}^{bI}v^2b^2)\chi_{II}/\alpha b^2 \, .
\end{aligned}
 \label{f7g7}
\end{equation} These equations together with Eq.~(\ref{point}) represent the list of
symmetry operators required for construction of analytical solutions.

\subsection{Invariant solutions} Let us now find the desired analytical solutions.
Equations (\ref{g})-(\ref{f}) with expressions substituted from Eq.~(\ref{f5g5}),
(\ref{f7g7}) represent a system of partial differential equations \begin{eqnarray}
 2bv\tau_{II} + \left( 2 - {\rm e}^{bI-1} - {v^2 b^2 {\rm
 e}^{bI}\over 2\alpha} \right) \chi_{II} + b\left( 1 - {\rm e}^{bI-1}
 - {v^2 b^2 {\rm e}^{bI}\over 2 \alpha} \right)
 \chi_{I} = 0\,,\label{1}\\
 -{2vb{\rm e}^{bI}\over \alpha}\chi_{II} - {vb^2{\rm e}^{bI}\over
 \alpha}\chi_I + \left( 2 - {\rm e}^{bI-1} - {v^2b^2 {\rm
 e}^{bI}\over 2 \alpha} \right) \tau_{II} - b\tau_I = 0\,.\label{2}
\end{eqnarray}

The first integral to the equation (\ref{1}) can be easily found: \begin{eqnarray}
2bv\tau_I + \left( 2 - {\rm e}^{bI-1} - {v^2b^2{\rm e}^{bI}\over 2\alpha} \right)
\chi_I + b\chi = J(v).\label{Int} \end{eqnarray} $J(v)$ in the above formula should be
found from the comparison with Eq.~(\ref{2}). Differentiating Eq.~(\ref{Int}) with
respect to $v$, taking Eqs. (\ref{BVP}) into account and comparing obtained expression
with Eq. (\ref{2}), one can see that $J(v)$ should be a constant. In view of a
symmetric initial intensity distribution with respect to $x \to -x$ reflections we are
bound to choose $J=0$. Then, substituting $\tau_I = -\chi_v$ into Eq.~(\ref{Int}), we
arrive at the following first order partial differential equation \begin{eqnarray}
-2bv\chi_v + \left(2 - {\rm e}^{bI-1} - {v^2b^2{\rm e}^{bI}\over 2\alpha} \right)\chi_I
+ b\chi = 0,\label{3} \end{eqnarray} which can be integrated with a standard
technique.

Integration of Eq.~(\ref{3}) gives two first integrals, \begin{eqnarray} J^1 =
\frac{-\chi^2}{v},\hspace{1cm} \quad J^2 = \frac{-1}{v} \left( 2 {\rm e}^{1-bI} - 1 +
\frac{b^2 {\rm e}}{2\alpha} v^2 \right). \label{firstint} \end{eqnarray} Here and in
what follows one has to keep in mind that a negative value of $v$ corresponds to the
focusing beam for the positive values of $x$.

Now we are in a position to find a particular solution $\chi(v,I)$ for the
Eqs.~(\ref{BVP}) satisfying the boundary conditions Eq.~(\ref{PS}). Let us first notice
that from the system of equations (\ref{BVP}) a linear second order partial
differential equation \begin{eqnarray} \alpha\chi_{vv} + ({\rm e}^{bI}\chi_I)_I =
0,\label{SecOrEq} \end{eqnarray} can be derived. Based on the result obtained in
Eqs.~(\ref{firstint}), one can search for the solution to Eq.~(\ref{SecOrEq}) based on
the following Ansatz \begin{equation} \chi^2 = -v Q^2(J^2),\label{sol1} \end{equation}
Substituting Eq.~(\ref{sol1}) into Eq.~(\ref{SecOrEq}) after some calculations we get:
\begin{eqnarray}
 Q_{\mu\mu} - Q/4=0,\label{so2}
\end{eqnarray} where $\mu = {\rm arcsinh}(J^2/\sqrt{2b^2{\rm e}/\alpha})$. Equation
(\ref{so2}) has an evident general solution \begin{eqnarray} Q = C_1{\rm e}^{-\mu/2} +
C_2{\rm e}^{\mu/2},\nonumber \end{eqnarray} where $C_1$ and $C_2$ are constants which
should be found from the boundary conditions.

Taking the Eq.~(\ref{sol1}) into account we obtain the expression for $\chi$:
\begin{eqnarray}
 \chi = (-v)^{1/2}\left[ C_1{\rm e}^{-\mu/2} + C_2{\rm e}^{\mu/2}
 \right]\, ,\label{chi}
\end{eqnarray} where ${\rm e}^{\mu/2}$ is to be found from the equation:
\begin{eqnarray} \quad \sinh \mu & = & \sqrt{\frac{\alpha}{2b^2{\rm e}}}\frac{-1}{v}
\left( 2 {\rm e}^{1-bI} - 1 + \frac{b^2 {\rm e}}{2\alpha} v^2 \right) \, .
\label{sol-h} \end{eqnarray}

Now we can express ${\rm e}^{\mu}$ from Eq.~(\ref{sol-h}) \begin{eqnarray} {\rm
e}^{\mu} = K/2\pm \sqrt{K^2/4+1} \, ,\label{e_mu} \end{eqnarray} where \begin{eqnarray}
K\equiv \sqrt{\frac{2\alpha}{b^2{\rm e}}} \frac{-1}{v} \left( 2 {\rm e}^{1-bI} - 1 +
\frac{b^2 {\rm e}}{2\alpha} v^2 \right). \nonumber \end{eqnarray} Summarizing, the
following solutions to the system of equations (\ref{BVP}) is obtained:
\begin{eqnarray} \chi = {1\over \sqrt{2}}\sqrt{ \left( 2 {\rm e}^{1-bI} - 1 + {b^2 {\rm
e} v^2 \over 2\alpha} \right) + \sqrt{ \left( 2 {\rm e}^{1-bI} - 1 + {b^2 {\rm e}v^2
\over 2\alpha} \right)^2 + { 2b^2{\rm e}v^2\over \alpha }} }. \label{Solution1}
\end{eqnarray}

In order to find the second function $\tau(I,v)$, we shall integrate the original
equation (\ref{BVP}) keeping the result (\ref{Solution1}) in mind. From
Eqs.~(\ref{BVP}) we have \begin{equation} \tau = \frac{1}{\alpha}\int\limits_{0}^{v}
{\rm d} v {\rm e}^{bI} \chi_I\,. \label{sol-t} \end{equation}

For the sake of convenience, let us introduce a new variable $\xi$: \begin{eqnarray}
\cosh (\xi)\equiv { 2 {\rm e}^{1-bI} + {b^2 {\rm e}v^2 \over 2\alpha} + 1 \over \sqrt{
8{\rm e}^{1-bI}} }. \end{eqnarray} Then \begin{eqnarray} \chi_I = {b\over 4
\sqrt{\Theta}} {{\rm e}^{1-bI\over 2} - \sqrt{2}{\rm e}^{\xi+1-bI} \over \sinh
(\xi)}.\nonumber \end{eqnarray} The expression for $\tau$ becomes: \begin{eqnarray}
\tau = {1\over b}\int_{\xi_0}^{\xi}{(a{\rm e}^{\xi}-1)^{1/2}\over -v}d\xi =
\left.\sqrt{{\rm e}\over 2 \alpha} \ln\left( {2{\rm e}^{\xi}\over a} - 1 +
\sqrt{\left({2{\rm e}^{\xi}\over a} - 1 \right)^2 - 1}\right)
\right\vert_{\xi_0}^{\xi},\nonumber \end{eqnarray} where $a\equiv\sqrt{2}\exp ((1-bI)/
2))$. Taking the boundary conditions into account, the final solution reads:
\begin{eqnarray} \tau = {\sqrt{{2 \rm e}\over \alpha}} \ln \left( {\rm
e}^{(bI-1)/2}\sqrt{\frac{(\chi^2+1)}{2}} + \sqrt{ {\rm e}^{bI-1}\frac{(\chi^2+1)}{2}
-1} \right).\label{Solution2} \end{eqnarray}

\begin{figure}
 \centering
 \includegraphics[angle=0,width=9cm]{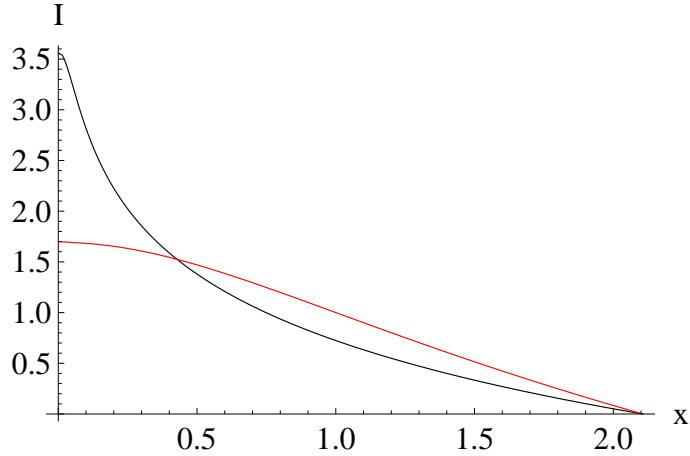} \caption{\label{I_0}
 Beam profile at different propagation distances. $\alpha=3,$ $b=1$.
 Red curve - $z=0,$ black curve - $z=0.8$.}
\end{figure}

After direct substitution of Eqs.~(\ref{Solution1}), (\ref{Solution2}) into
Eqs.~(\ref{BVP}) and a tedious calculation it is possible to verify that the obtained
functions $\chi(I,v)$ and $\tau(I,v)$ are indeed exact analytical solutions for the
formulated boundary value problem. In Fig.~\ref{I_0} we plot the intensity beam
distribution at different propagation distances calculated on the basis of the found
solutions Eqs.~(\ref{Solution1}), (\ref{Solution2}). 

Let us now examine the obtained result a little more closely. Firstly let us find the
total radius of the beam as a function of propagation distance $z$. For this goal, it
is necessary to determine where the intensity implicitly given by
Eqs.~(\ref{Solution1}), (\ref{Solution2}) intersects the surface of $x=0$. Putting
$I=0$ in Eq.~(\ref{Solution2}), we find $\chi|_{I=0}=\pm \sqrt{2{\rm e} - 1}$.
Substituting this number into Eq.~(\ref{Solution1}) one can deduce that $v = 0$ and,
consequently, $x = \pm \sqrt{2{\rm e} - 1}$. Thus, the phase gradient at the beam edge
is equal to zero, the total radius of the beam is a constant and does not depend on the
propagation length. By the numerical integration of the solutions
Eqs.~(\ref{Solution1}), (\ref{Solution2}) one can verify their consistence with energy
conservation, i.e. $\int I(x, z) dx$ from $x = \sqrt{2{\rm e}-1}$ to $x = -\sqrt{2{\rm
e}-1}$ is a constant.

The fact that the total radius of the beam for the case under consideration remains
constant and does not depend on $z$ is a new one and completely different from all
exact analytical results obtained so far. It was demonstrated earlier
\cite{Akhmanov,Kovalev1+1} that for the Kerr nonlinearity the total beam radius
decreases upon beam propagation. In the present case, the beam shape and peak intensity
are thus the only parameters depending on the propagation distance.

\begin{figure}
 \centering
\includegraphics[angle=0,width=9cm]{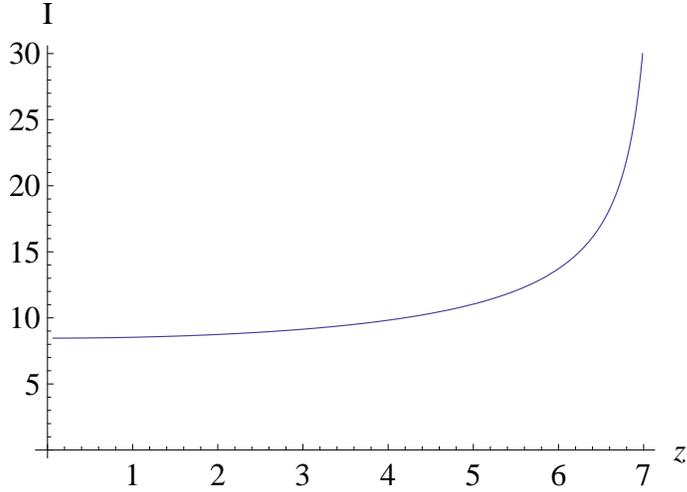} \caption{\label{Onaxis} On-axial intensity
distribution versus the propagation distance. } \end{figure}

Evolution of the beam peak intensity, which for the symmetry reasons has to be situated
on the beam axis, can be easily found from Eq.~(\ref{Solution2}). Putting $x=0$, $v=0$
we have \begin{equation}
 \frac{z}{b}\sqrt{\frac{2\alpha}{\rm e}} = \frac{2}{bI} \ln \left( (1/\sqrt{2})
 \, {\rm e}^{(bI-1)/2} + \sqrt{ (1/2) \, {\rm e}^{bI-1} -1} \right) .
 \label{onAxis}
\end{equation} On-axial intensity distribution versus the propagation distance is
presented in Fig.~\ref{Onaxis} for $\alpha=0.001,$ $b=0.2.$ We see that the intensity
monotonically increases and tends to infinity for $z$ approaching a critical value
denoted as a self-focusing position $z_{\rm sf}$. Its exact value can be found from
direct analysis of the Eq.~(\ref{onAxis}). Considering Eq.~(\ref{onAxis}) in the limit
$I \to \infty$ one obtains \begin{eqnarray} z_{\rm sf} = b\sqrt{{\rm e}\over
2\alpha}.\label{zsfExact} \end{eqnarray} A detailed investigation of the Eqs.
(\ref{Solution1}), (\ref{Solution2}) shows that the solutions exhibit no singularities
before this point.

\section{Approximate solutions to the eikonal equation in (1+1) dimensions with
arbitrary refractive index}

Because it is not possible to construct an exact analytical solution for every desired
form of the refractive index and the boundary conditions, let us now present here a
possible way to obtain approximate analytical solutions. We shall start from the
Eqs.~(\ref{hod1}). Under certain conditions (see e.g. \cite{TatGar2}), $\alpha$ can be
considered as a small parameter. Due to its smallness, we will search for an
approximate symmetries group operator: \begin{eqnarray}
X=p\partial_{\tau}+q\partial_{\chi},\nonumber \end{eqnarray} with coordinates in form
of a power series in $\alpha$: \begin{eqnarray}
 p=\sum \alpha^i p^i\,,\qquad q=\sum \alpha^i q^i\,,\qquad
  i = 0,\ldots ,\infty\,.
 \label{fgapprox}
\end{eqnarray} In case of Eqs.~(\ref{hod1}), the determining equations read:
\begin{equation}
  \left( D_w^0 + \alpha D_w^1 \right) p - \frac{I}{\varphi(I)} D_I q=0\,,\quad
  \left( D_w^0 + \alpha D_w^1 \right) q + \alpha D_I p=0 \,,
 \label{Det2}
\end{equation} where $w=v/\alpha$ and \[
 \begin{aligned} &
 D_w^0 = \partial_w + \sum_{s=0}^{\infty}\left((I/\varphi)\chi_I \right)_{I^{s}}
 \partial_{\tau_{I}^s}\,, \quad
 D_w^1 = - \sum_{s=0}^{\infty} \tau_{I}^{s+1}
 \partial_{\tau_{I}^s}\,.
 \end{aligned}
 \]
Substituting Eqs.~(\ref{fgapprox}) into Eqs.~(\ref{Det2}) we arrive at a following
system of recurrent differential equations: \begin{eqnarray}
 && D_w^0 q^i + (1-\delta_{i,0}) \left( D_w^1 q^{i-1} + D_I (p^i) \right) = 0 \,,
 \label{gi}\\*
 && D_w^0 p^i + (1-\delta_{i,0}) D_w^1 p^{i-1}-\frac{I}{\varphi}D_I (q^i)=0\,.
 \label{fi}
\end{eqnarray} The system (\ref{gi})-(\ref{fi}) can be solved sequentially starting
from a given $g^0$. Thus, integration of Eqs.~(\ref{gi})-(\ref{fi}) gives
\begin{equation} \begin{aligned}
 p^{i}=\int
 dw  &
 \left\{\sum_{s=0}^{\infty}(1-\delta_{i,0})\tau_I^{s+1}\partial_{\chi_I^s}p^{i-1}
 \right.
  \\
  &  \left.  +\frac{I}{\varphi}\left[ \partial_I
 +\sum_{s=0}^{\infty}\left(\tau_I^{s+1}\partial_{\tau_I^s}+\chi_I^{s+1}\partial_{\chi_I^s}
 \right)
 \right]q^i \right\}+P^i\,, \\
 q^{i}=(1-\delta_{i,0})\int
 dw & \left\{\sum_{s=0}^{\infty}\tau_I^{s+1}\partial_{\chi_I^s}q^{i-1} \right. \\ &
 \left.
 -\left[
  \partial_I +\sum_{s=0}^{\infty}\left(\tau_I^{s+1}
  \partial_{\tau_I^s}+\chi_I^{s+1}\partial_{\chi_I^s}\right)
 \right]p^{i-1} \right\}+Q^i \,.
 \end{aligned}
\end{equation} Here $P^i$ and $Q^i$ are arbitrary functions of invariants \[ I, \quad
\chi_I^s,  \quad \mbox{and} \quad
 \tilde{\tau_I^s}=\tau_I^s-w\sum_{p=0}^s C^s_p (\partial^p(I/\varphi)/\partial
 I^p) \chi_I^{s-p+1}, \quad i\geq 1 \,,
\] and $C^s_p$ is the binomial coefficient (see Ref.~\cite{Kovalev1+1}).

Let us now put \begin{eqnarray}
 q^0=0, \quad p^0=1+2I \chi\chi_I/\varphi \,.\label{f0app}
\end{eqnarray} This choice of $p^0$ corresponds to a special form of initial light beam
\[ \int_0^{I_0} (\varphi/I)\, dI=-x^2 \,. \] For nonlinearities $\varphi(I)$ relevant
to particular physical situations, such form of initial intensity distribution is very
similar to the Gaussian profile. It should be stressed that, in general, there are no
restrictions on the initial intensity distribution $I=I_0(x)$, the present choice is
made only for the sake of further simplicity. \par

Substituting this function into equation for $g^1$ we get \begin{equation}
 q^1=wD_I(2I\chi\chi_I/\varphi)
 +G^1 = {\varphi(\tilde{\tau}-\tau)\over I
 \chi_I}D_I(2I\chi\chi_I/\varphi) + G^1=-2(\tau \chi)_I\,,\label{g1app}
\end{equation} where the term with ${\tilde \tau}$ was included into a new arbitrary
function $\tilde{G^1}$, which was later put equal to zero.

Using this result, we can calculate $p^1$ in a similar way. We get \begin{equation}
 p^1=-{I\over \varphi}\left( \tau^2{\varphi \over I}
 \right)_I \,. \label{f1app}
\end{equation} Finally, up to the first order of $\alpha$, the coordinates of the
approximate symmetries group generator read \begin{equation}
 q = -2\alpha (\tau\chi)_I\,, \quad
 p =1+2I\chi\chi_I/\varphi - \alpha \frac{I}{\varphi}
 \left( \frac{\tau^2\varphi}{I} \right)_I \,.
\end{equation} Taking Eqs.~(\ref{hod1}) into account, we rewrite the desired symmetry
group operator as follows \begin{eqnarray} R=\left( 1-\alpha \tau^2{(\varphi/I)_I\over
\varphi/I} \right)\partial_{\tau}-2\chi \partial_w + 2 \alpha \tau \partial_I
\,.\label{RR} \end{eqnarray} Integration of the Lie equations corresponding to the
point symmetry operator (\ref{RR}) gives us the three first integrals whose particular
form depends on the choice of $\varphi=\varphi(I)$. We write these integrals by
introducing the new function $\phi(I)$, such that $\phi_I = \varphi/I$:
\begin{equation}
 J_1=\chi \,, \quad   J_2=\alpha \tau^2 \phi_I - \phi \,, \quad
 J_3 = w+ \frac{\chi}{\sqrt{\alpha}} \int
  \frac{ {\rm d} \phi }{\sqrt{\phi_I} \sqrt{\phi + J_2} } \,. \label{1ints}
\end{equation} Now, the solution is a function of these first integrals fulfilling the
boundary conditions. This means that values of $J_i$ are taken from the boundary
conditions when $\tau \to 0$ and $\chi \to x^{\prime}$. Hence we rewrite (\ref{1ints})
in the following form \begin{equation} \begin{aligned} &
 x^{\prime} = \chi \,, \qquad   \phi(I_0(x^\prime)) = - \alpha z^2 I^2 \phi_I + \phi
 \,, \\ &
 x^{\prime} \sqrt{\alpha} \int\limits^{\phi(I_0(x^\prime)) }   \frac{ {\rm d} \phi
 }{\sqrt{\phi_I} \sqrt{\phi + J_2} } =
 v + (x-vz) \sqrt{\alpha} \int\limits^{\phi}
 \frac{ {\rm d} \phi }{\sqrt{\phi_I} \sqrt{\phi + J_2} } \,.
 \end{aligned}
 \label{solapr}
\end{equation} In the particular case of Kerr nonlinearity $\varphi=1$, we arrive at
the solution previously obtained in Ref. \cite{Kovalev1+1}.

Let us now compare the approximate solutions with the exact solution constructed in the
previous section. In order to satisfy the boundary condition Eq.~(\ref{PS}), $p^0$ is
taken in the form \begin{eqnarray}
 p^0=2b{\rm e}+2I\chi \chi_I/\varphi \,,\nonumber
\end{eqnarray} Then the generator (\ref{RR}) reduces to: \begin{eqnarray} R=(2b{\rm
e}+\alpha \tau^2)\partial_{\tau}-2\chi \partial_w +2
 \alpha \tau \partial_I \,,\nonumber
\end{eqnarray} and yields an approximate solution: \begin{equation}
 \begin{aligned}&
 \left( \left( x-vz \right)^2+1 \right) {\rm e}^{bI}=\alpha I^2 z^2+2{\rm e} \,,  \\ &
 v = -\sqrt{\frac{2 \alpha }{{\rm e}}} \frac{1}{b}(x-vz) \,
 {\rm arctg}\left( I z \sqrt{\frac{\alpha}{2 {\rm e}}}\right) \,.
 \end{aligned}
\label{approx} \end{equation} On-axial intensity distribution is given by expression
${\rm e}^{bI} - 2{\rm e}=\alpha I^2 z^2$, which exhibits no singularities. The on-axial
intensity monotonically increases upon propagation. However, the function $v(x,t)$ can
exhibit singularities which, due to the symmetry of the problem, are expected to be on
the beam axis. Let us investigate this behavior more closely. In the vicinity of the
beam axis, $v_x$ can be approximated as
 \[ v_x=-
 \sqrt{\frac{2\alpha}{{\rm e}}} \frac{1}{b} \, {\rm arctg}\left( Iz
 \sqrt{\frac{\alpha}{2 {\rm e}}} \right)
 \left(
 1- z \sqrt{\frac{2 \alpha}{\rm e}} \frac{1}{b} \,{\rm arctg} \left( Iz
 \sqrt{\frac{\alpha}{2 {\rm e}}}\right) \right)^{-1} \,.
 \]
Then, the position of a singularity at the beam axis (what corresponds to the rays
intersection) can be found from the system of equations: \begin{equation}
 1- z_{sf} \sqrt{\frac{2 \alpha}{\rm e}} \frac{1}{b} \,{\rm arctg}
 \left( Iz_{sf} \sqrt{\frac{\alpha}{2 {\rm e}}}\right) = 0 \,, \qquad
 \alpha z_{sf}^2 I^2 =  - 2{\rm e} + {\rm e}^{bI}\,.
 \label{zsfappr}
\end{equation} The numerical solution of Eqs.~(\ref{zsfappr}) yields $z_{\rm sf}
\approx 1.03 \, b \sqrt{{\rm e}/ 2 \alpha}$. As one sees, this result is very similar
to Eq. (\ref{zsfExact}) obtained from the exact solution. The similar tendency has been observed in Ref. \cite{Kov-Nl} for the Kerr refractive index: an approximate solution provides a longer self-focusing distance in comparison to exact one.

\section{Approximate solution to the Schr\"odinger equations with arbitrary refractive
index in (1+2) dimensions}

In this section we shall turn to the construction of approximate solutions for the NLSE
(\ref{NLSE}) in (1+2) dimensions in media with arbitrary nonlinearity. Such a
mathematical model describes the propagation of a continuum wave beam in cylindrical
geometry and has great number of application to particular physical situations (see
e.g. \cite{Boyd,CPhysRep,BergeRPP}). Let us begin with equations (\ref{BE1a}),
(\ref{BE1b}) supplemented by the boundary condition
\begin{eqnarray}
v(0,x)=0 \,,
\qquad I(0,x)=N(x) \,,
\end{eqnarray}
which corresponds to a collimated beam with
arbitrary initial intensity distribution.

As usual, we start from construction of the Lie-B\"acklund symmetry operator of the
form
\begin{eqnarray} X= \kappa \partial_v + \lambda \partial_I\,.\nonumber
\end{eqnarray}
The determining equations read:
\begin{equation}
 \begin{aligned}
 D_z \kappa & + v D_x \kappa  + \kappa v_1-\alpha \varphi D_x \lambda -\alpha \varphi_I
 I_1
 \lambda \\
  & - \beta\left( B_I \lambda + B_{I_1} D_x \lambda + B_{I_{2}}D_{x}^2 \lambda
  + B_{I_{3}} D_{x}^3 \lambda \right)=0, \\
  D_z \lambda & + I D_x \kappa  + v D_x \lambda  + \lambda v_1 + \kappa I_1
  + \frac{v \lambda}{x}+ \frac{\kappa I }{x}=0 \,,
 \end{aligned}
 \label{det12}
\end{equation} where
\begin{equation}
 \begin{aligned} &
 B = D_x \left( {D_x(xD_x\sqrt{I})\over x\sqrt{I}} \right), \quad
  v_s \equiv \frac{\partial^s v}{\partial x^s } \,,
 \quad   I_s \equiv \frac{\partial^s I}{\partial x^s } \,,  \\ &
 D_x = \partial_x + \sum_{s=0}^{\infty}(v_{s+1}\partial_{v_s}+I_{s+1}\partial_{I_s}),
 \label{dx}
\end{aligned}
\end{equation}
and $D_z$ we present as $D_z=D_z^0+D_z^1,$ where
\[
 \begin{aligned} &
 D_z^0 = \partial_z - \sum_{s=0}^{\infty}\left(
 D_x^s(vv_1)\partial_{v_s}+\left[ D_x^{s+1}(Iv) +D_x^s\left(\frac{Iv}{x} \right)
 \right] \partial_{I_s}
 \right)\,, \\ &
 D_z^1 = \sum_{s=0}^{\infty} \left[ D_x^s(\alpha \varphi I_1 + \beta B) \right]
 \partial_{v_s} \,.
 \end{aligned}
\]

Since $\alpha$ and $\beta$ in Eqs.~(\ref{BE1}) can be considered as small
parameters, we will search for $\kappa$ and $\lambda$ in form of a series expansion in
powers of $\alpha$ and $\beta$
\begin{eqnarray}
 \kappa =\sum_{i,j=0}^{\infty}\alpha^i \beta^j \kappa^{(i,j)}, \qquad
 \lambda=\sum_{i,j=0}^{\infty}\alpha^i \beta^j \lambda^{(i,j)} \,.
 \label{fijgij}
\end{eqnarray} and restrict ourselves only to the first order corrections
\begin{eqnarray}
 \kappa =\kappa ^0+\kappa ^1+O(\alpha^2,\beta^2,\alpha\beta)\,, \qquad
 \lambda=\lambda^0+\lambda^1+O(\alpha^2,\beta^2,\alpha\beta)\,,
 \label{fijgij00}
\end{eqnarray} where $\kappa^0 \equiv \kappa^{(0,0)}$, $\lambda^0 \equiv
\lambda^{(0,0)}$, $\kappa^1 \equiv \alpha \kappa^{(1,0)} + \beta \kappa^{(0,1)}$, and
$\lambda^{1} \equiv \alpha \lambda^{(1,0)} + \beta\lambda^{(0,1)}$. Let us now write
down the determining equations keeping only the linear terms with respect to $\alpha$
and $\beta$. We get \begin{equation}
 \begin{aligned}
  M_0 \kappa^0 & = 0 \,, \qquad M_1 \lambda^0 + M_2 \kappa^0=0 \,, \\
  M_0 \kappa^1 & + D_z^1 \kappa^0-\alpha D_x(\varphi \lambda^0) \\
  & - \beta \left[ B_I \lambda^0+B_{I_x}D_x \lambda^0+B_{I_{xx}}D^2_x \lambda^0
  + B_{I_{xxx}}D^3_x \lambda^0 \right] =0 \,, \\
 M_1 \lambda^1 & + D_z^1 \lambda^0+M_2 \kappa^1 = 0 \,,
 \end{aligned}
 \label{D1}
\end{equation} where \begin{equation}
 \begin{aligned} &
 M_0 = D_z^0+vD_x+v_1 \,, \\ &
 M_1 = D_z^0+vD_x+v_1+ v / x \,, \\ &
 M_2 = ID_x+I_1+ I / x \,.
 \end{aligned}
 \label{L0}
\end{equation} Let us now following to Ref. \cite{Kovalev1+2} put \begin{equation}
 \kappa ^0= \frac{1}{2} D_x(v^2)\,,\qquad \lambda^0= \frac{1}{x} D_x(Ivx)\,.
 \label{f0g00}
\end{equation} Evidently this choice satisfies the zero-order Eqs.~(\ref{D1}) and the
invariance conditions: $\kappa^0=0$, $\lambda^0=0$ at the boundary. Then $\kappa^1$ can
be found from the first of the first-order equations in Eq.~(\ref{D1}) that is
rewritten as \begin{equation}
 M_0 \left( \kappa ^1 + \alpha \varphi I_1 + \beta D_x \left(
 \frac{D_x(xD_x\sqrt{I})}{x\sqrt{I}} \right)
 \right)=0.
 \label{L0f1}
\end{equation} The solution of this equation is expressed in terms of invariants of the
operator $M_0$, \begin{eqnarray}
 \kappa^1 =  D_x \left( S(\chi) - \alpha \Phi - \beta \frac{D_x(xD_x\sqrt{I})}{
 x\sqrt{I}}  \right) \,,
 \label{f13dapp}
\end{eqnarray} where \begin{equation}
 \varphi = \partial_I \Phi \,, \quad S(\chi) \equiv \alpha \Phi(N(\chi))
 + \beta \frac{(\chi (\sqrt{N(\chi)})_\chi)_\chi}{ \chi \sqrt{N(\chi)}} \,.
 \label{S}
\end{equation} Inserting this result into the second of the first-order equations in
Eq.~(\ref{D1}) we get the equation for the function $q^1$ \begin{eqnarray}
 M_1 \lambda^1 + {D_x\over x}\left( Iz D_x S(\chi) \right)=0 \,.\nonumber
\end{eqnarray} It is easy to show by direct substitution that the formula above can be
rewritten as \[
 M_0 (x \lambda^1) + D_x \left( Iz D_x S(\chi) \right) = 0 \,.
\] This equation can be integrated in a same as (\ref{L0f1}). Then one gets \[
 x \lambda^1=-D_x(xIz S_{\chi}). \nonumber
\] Finally, up to the first order in the small parameters, the Lie-B\"acklund symmetry
operators in the canonical form read:
 \begin{eqnarray}
 & & \kappa = v v_1 + D_x \left( S(\chi) - \alpha \Phi - \beta
 \frac{D_x(xD_x\sqrt{I})}{ x\sqrt{I}}  \right) \,,
 \label{f3Dapp} \\
 & & \lambda = v \left( I_1 + \frac{I}{x} \right) + I v_x - z \left[ I(1-z v_1) S_{\chi
 \chi} +
 \left( I_1 + \frac{I}{x} \right) S_{\chi} \right] \,. \label{g3Dapp}
\end{eqnarray} We notice that based on (\ref{BE1a}), (\ref{BE1b}) Eq.~(\ref{f3Dapp})
can be rewritten as follows: \[
 \kappa = v_z-(1-v_xz)S_{\chi} \,.
\]

Together with Eq.~(\ref{f3Dapp}), the equation above lead to two relations:
\begin{eqnarray}
 & & v = zS_{\chi} \,, \label{Const1} \\
 & & v_z = (1-v_xz)S_{\chi} \,, \label{Const2}
\end{eqnarray} that have to be fulfilled in order to preserve the invariance
requirement $\kappa=0$, $\lambda=0$.

Now, keeping the relation between the canonical form of the symmetries operator and the
point symmetries group operator \cite{Ibragimov} in mind we can write down the group
symmetry operator: \begin{equation}
 \begin{aligned}
 R & = \left( 1+z^2S_{\chi\chi}
 \right)\partial_z+S_{\chi}\partial_v+(zS_{\chi}+
 vz^2S_{\chi\chi})\partial_x  \\
 & - I z \left( \left(1+{vz\over x}\right)S_{\chi\chi}+
 {1\over x}S_{\chi}\right)\partial_I \,.
\end{aligned}
 \label{R3D}
\end{equation} Operator Eq.~(\ref{R3D}) is similar to the one obtained previously in
Ref.~\cite{Kovalev1+1} for a collimated beam with the exception that the $S(\chi)$ now
contains an arbitrary function $\Phi$. The generator (\ref{R3D}) yields a system of
characteristic equations: \begin{eqnarray} {dz\over 1+z^2S_{\chi\chi}}={dv\over
S_{\chi}}={d\chi \over -v}={d \ln(Ix)\over -zS_{\chi\chi}}.\label{Char} \end{eqnarray}
This system of equations can be easily integrated after taking into account
Eq.~(\ref{Const1}). The second and third equations of (\ref{Char}) give
\begin{eqnarray}
 S+{S_{\chi}^2z^2\over 2}=S(\mu)\,,\label{Smu}
\end{eqnarray} where $\mu$ corresponds to the value of $\chi$ at the boundary. The
third and the fourth of Eqs.~(\ref{Char}) yield another invariant $Ix/S_{\chi}$, what
can also be rewritten as \begin{eqnarray}
 I=N(\mu){\chi\over x}{S_{\chi^2}\over S_{\mu^2}} \,, \label{IN}
\end{eqnarray} where $N(\chi)$ is an initial intensity profile. \par

From the first and the second of Eq.~(\ref{Char}) we have
$v_z=S_{\chi}/(1+z^2S_{\chi\chi})$. Taking Eq.~\ref{Const2}) and $D_x S(\chi) =(1-z
v_x) S_\chi$ into account and using $ 2 \chi S_{\chi}=S_{\chi^2}$, we finally arrive at
a relation between $x$ and $\chi$: \begin{eqnarray} x=\chi (1+2z^2 S_{\chi^2}) \,.
\label{x-chi} \end{eqnarray} Summarizing, the solutions are presented by the equations:
\begin{eqnarray}
 v(x,z)={x-\chi\over z}\,, \qquad I(x,z)=N(\mu) \frac{\chi}{x}
 \frac{S_{\chi^2}}{S_{\mu^2}} \,,
 \label{Sol3d}
\end{eqnarray} where $\chi$ and $\mu$ are defined as functions of $z$ and $x$ via
relations \begin{eqnarray} x=\chi \left( 1+2z^2 S_{\chi^2} \right) \,,
 \qquad S(\mu)=S(\chi)+\frac{S_{\chi}^2 z^2 }{2} \,,
 \label{Sol3dDop}
\end{eqnarray} These solutions describe the evolution of a collimated continuous wave
laser beam with arbitrary initial intensity distribution in media with arbitrary
nonlinear response. In case of Kerr refractive index, these solutions were investigated
in details in Refs. \cite{Kovalev1+2,KovalevPRA}. 
obtained solutions to more complicated forms of the refractive index. 

We shall consider refractive index of the form \begin{eqnarray} n=n_0+n_2I-\sigma_K
I^K.\label{nn2} \end{eqnarray} Here, the second term in the right hand side represents
the usual Kerr response, and the last term is responsible to the multiphoton ionization
of the media in case of sufficiently strong electric field; the $K$ then corresponds to
the number of photons required for a simultaneous ionization, and $\sigma_K$ to the
ionization rate. If at the entry plane of the nonlinear media the Gaussian intensity
distribution is fulfilled, one obtains \begin{eqnarray} S=\alpha \exp(-\chi^2)-\alpha
\gamma \exp(-K\chi^2)/K+\beta (\chi^2-2),\label{S2} \end{eqnarray} where $\gamma= K
\sigma_K I_0^{K-1}/n_2.$

Let us write the first of Eq. (\ref{Sol3d}) as follows \begin{eqnarray}
 v={2zS_{\chi^2}\over 1+2z^2S_{\chi^2}} \,, \label{v2}
\end{eqnarray} where $S_{\chi^2}=-\alpha {\rm e}^{-\chi^2}+\alpha\gamma {\rm
e}^{-K\chi^2}+\beta$. Here $v$ is a single-valued function of $x$ if the function
$x=\chi(x)$ can be determined from the first of equations (\ref{Sol3dDop}) uniquely. In
order to find the region of multivaluedness let us investigate the function $Y\equiv
\chi (1+2z^2 S_{\chi^2})$. We find where its first and second derivatives with respect
to $\chi$ vanish.

From equation $Y_{\chi}=0$ we have \begin{eqnarray} 1+2z^2 S_{\eta}(\eta)+4z^2\eta
S_{\eta\eta}(\eta)=0 \,,\label{Yx} \end{eqnarray} where $\eta\equiv \chi^2.$ Equation
$Y_{\chi\chi}=0$ gives \begin{eqnarray}
 \chi \left( 3S_{\eta\eta}(\eta)+2\eta S_{\eta\eta\eta}(\eta) \right) = 0 \,.
 \label{Yxx}
\end{eqnarray} Evidently, this equation is fulfilled if $\chi=0$, or the expression in
the brackets is equal to zero. In the first case, we observe the singularity  at the
beam axis, in the second case, the singularity takes place at the point
$\eta=\eta_{cr}$ which should be found numerically from equation
$3S_{\eta\eta}(\eta)+2\eta S_{\eta\eta\eta}(\eta)=0$ for each particular form of
$S(\eta).$

In the first case, when $\chi=0$, expression (\ref{v2}) becomes singular on the beam
axis at the point \begin{eqnarray} z_{\rm sf}=1/\sqrt{2(\alpha(1-\gamma)
-\beta)}.\label{zsf3d} \end{eqnarray} The beam collapse at the beam axis occurs if
$\alpha > \alpha\gamma + \beta$. This result is very similar to the case of Kerr
nonlinearity considered in Refs. \cite{Kovalev1+2,KovalevPRA}, the difference only
coming from the presence of the $\gamma$ term under the square root. If $\alpha <
\alpha\gamma + \beta$, the beam collapse does not take place on the beam axis.

In the second case, if $\chi\neq 0$, position of the singularity can be found form the
magnitude of $\eta_{cr}$ which gives us $\chi=\pm \sqrt{\eta_{cr}}$ and
$S_{\eta}(\eta)|_{\eta=\eta_{cr}}$. Starting form Eq. (\ref{Yx}), we can find the
coordinates of the singularity position: \begin{eqnarray} z=\left.\sqrt{-1\over
2(S_{\eta}+2S_{\eta\eta})}\right\vert_{\eta=\eta_{cr}},\hspace{1cm} x=\left.\pm
2\sqrt{\eta}\left({S_{\eta\eta}\over S_{\eta}+2S_{\eta\eta}}\right)
\right\vert_{\eta=\eta_{cr}}.\label{xz} \end{eqnarray}

\begin{figure}[!h]
\begin{minipage}{0.49\textwidth} \includegraphics[angle=0,width=0.95\textwidth]
{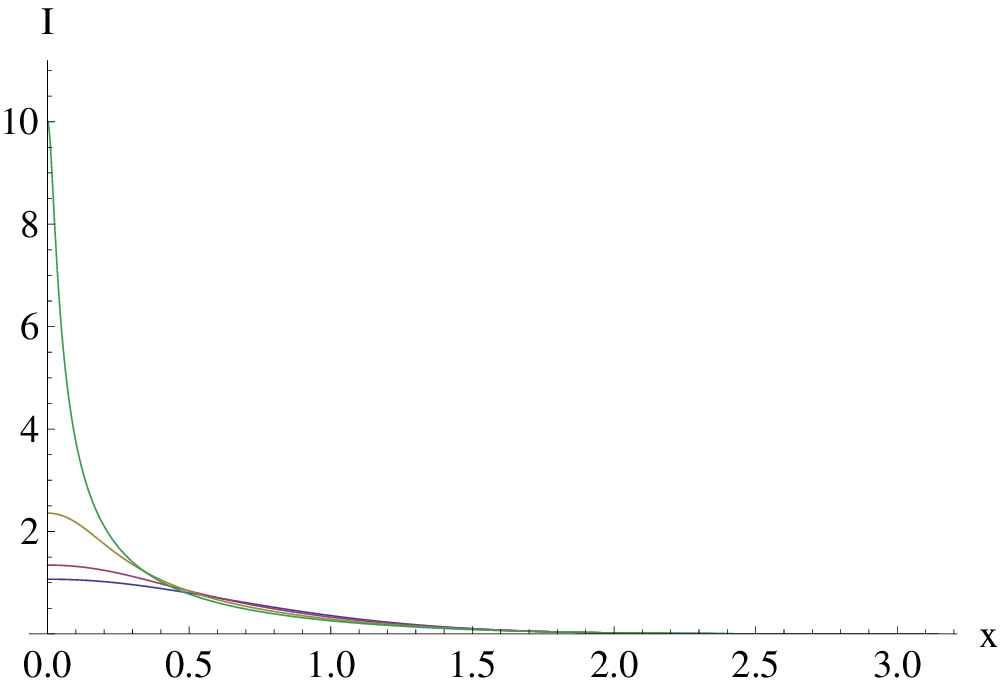} \end{minipage} \hfil \begin{minipage}{0.49\textwidth}
\includegraphics[angle=0,width=0.95\textwidth] {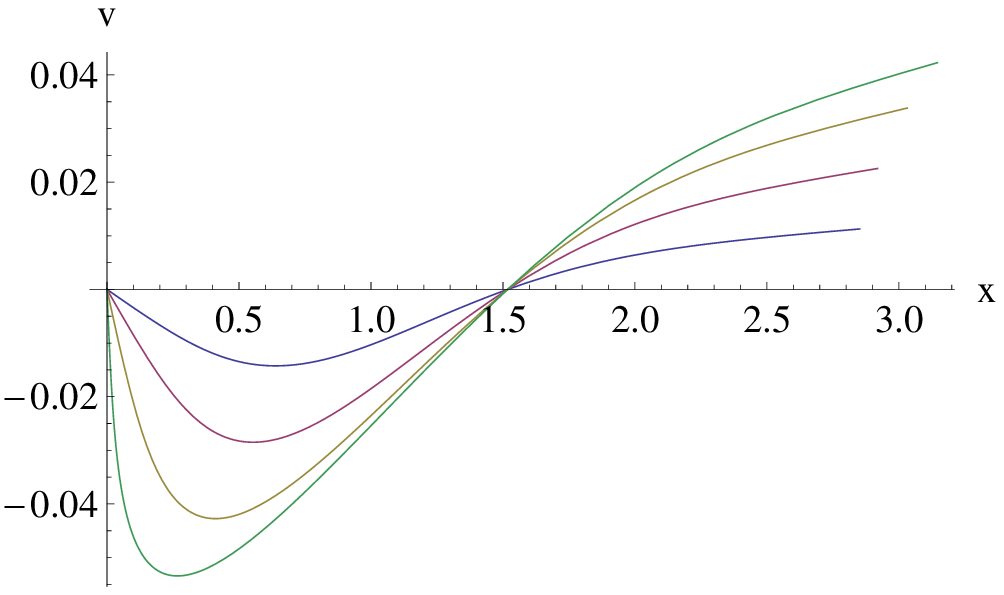} \end{minipage}
\caption{\label{ProfCollIv} The beam intensity (left panel) and the phase gradient
(right panel) distributions upon the transverse coordinate $x$ for $\alpha=0.01,$
$\beta=0.001,$ $\gamma=0.1$ and $K=6$. The transition near the beam axis $x\to 0$ from
the bottom to the top on the left panel and from the top to the bottom on the right
panel correspond to curves at different distances $z=2,$ $z=4,$ $z=6,$ $z=7.5$.}
\end{figure}

Let us now consider two particular choices of the parameters in the refractive index.
Let $\alpha=0.01,$ $\beta=0.001,$ $\gamma=0.1$, $K=6$. At the beam axis $x=0$, the
formula (\ref{zsf3d}) gives us the self-focusing position  $z_{\rm sf} \approx 7.9$.
Eq.~(\ref{Yxx}) has one solution $\eta \simeq 1.5$, but the corresponding magnitude of
the self-focusing distance defined by Eqs.~(\ref{xz}) is imaginary. This means that one
observes only one self-focusing position at the beam axis. The beam intensity, $I(x)$,
and the phase gradient, $v(x)$, as a function of $x$ at different propagation distances
are presented on Fig. \ref{ProfCollIv}. 

\begin{figure}[!h]
\begin{minipage}{0.49\textwidth} \includegraphics[angle=0,width=0.95\textwidth]{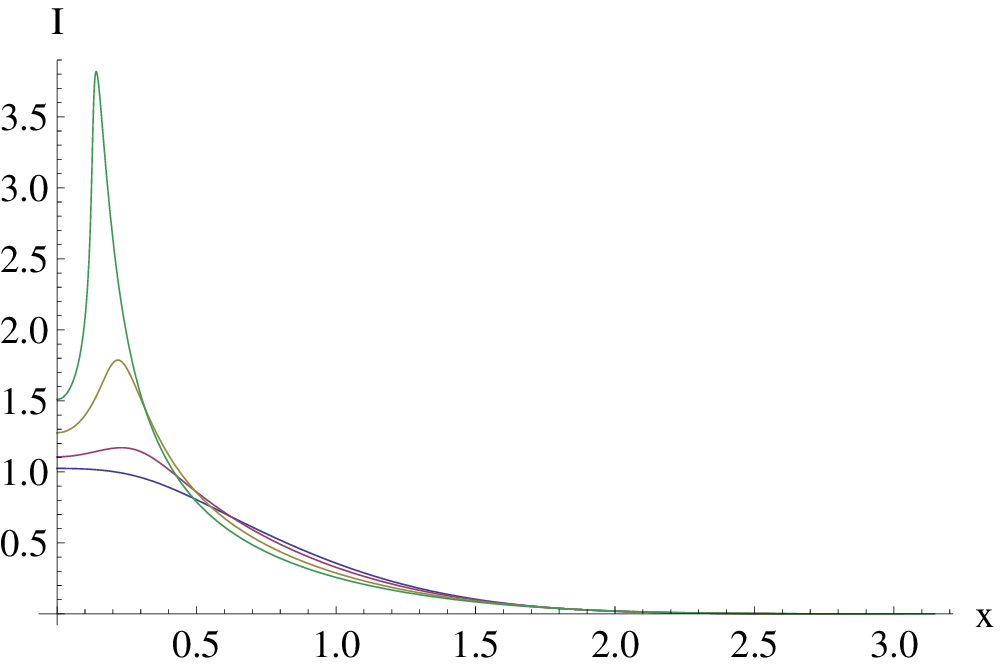}
\end{minipage} \hfil \begin{minipage}{0.49\textwidth}
\includegraphics[angle=0,width=0.95\textwidth]{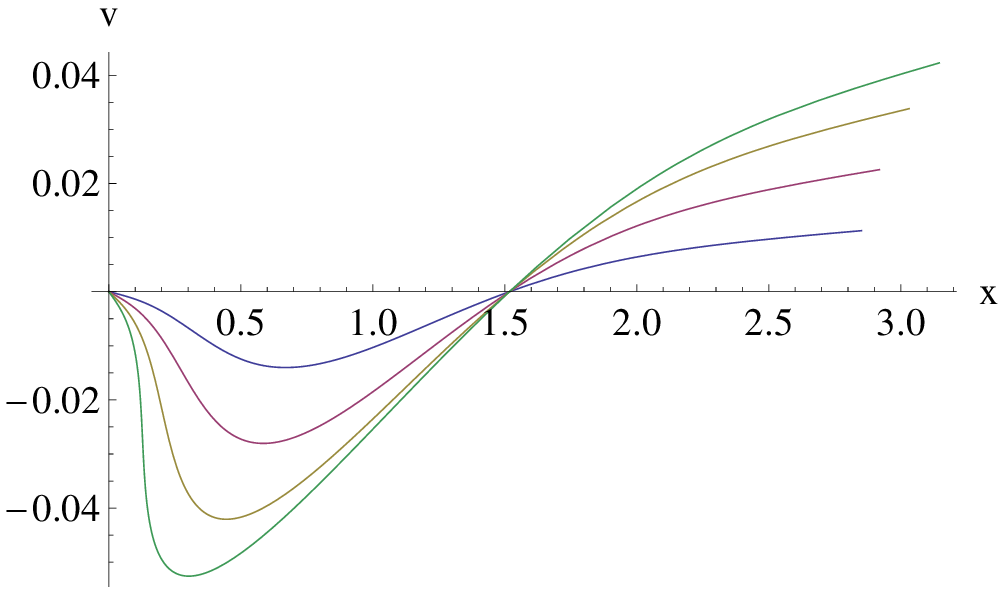} \end{minipage}
\caption{\label{ProfKolIv} The beam intensity (left panel) and the phase gradient
(right panel) distributions upon the transverse coordinate $x$ for $\alpha=0.01,$
$\beta=0.001,$ $\gamma=0.6$ and $K=8$. The transition near the beam axis $x\to 0$ from
the bottom to the top on the left panel and from the top to the bottom on the right
panel correspond to curves at different distances $z=2,$ $z=4,$ $z=6,$ $z=7.5$.}
\end{figure}

Let us now assume that $\alpha=0.01,$ $\beta=0.001,$ $\gamma=0.6$, $K=8$, then Eq.
(\ref{Yxx}) gives us two solutions $\eta_1=1.5$ and $\eta_2\simeq 0.11$. Similar to the
previous case, the first value of $\eta=\eta_1$ gives no singularity. However,
substituting $\eta_2$ into Eqs. (\ref{xz}), we get position where the beam collapse
take place $x\simeq0.1,\,z\simeq 8.$ Considering behavior of the solutions at the beam
axis, we see that the intensity increases, and the solution becomes singular at the
point $z_{cr}\simeq 13$, that already behind the distance at which the first
singularity appeared. The intensity and phase gradient at different propagation
distances are presented on the Fig.~\ref{ProfKolIv}.  Summarizing this part of the work, we can notice that for the form of the
refractive index Eq.~(\ref{nn2}) considered as an example in the present paper, several
pictures in the global behavior of the solutions can be distinguished: i) if the
solution of Eq.~(\ref{Yxx}) provide us with only imaginary values of $z_{\rm sf}$,
there is no beam collapse at all, ii) the singularity appears at the beam axis, ii) the
singularity appears around the beam axis at the circle with radius given by
Eqs.~(\ref{xz}).

\section{Conclusion} In the presented paper, making use of the Lie symmetry analysis,
we constructed exact and approximate analytical solutions for the problem of light
propagation in highly nonlinear media. For the first time exact analytical solutions
(\ref{Solution1}), (\ref{Solution2}) to the eikonal equations in (1+1) dimensions were
found with nonlinear refractive index being a saturated function of intensity. It was
shown that at a certain point at the beam axis Eq. (\ref{zsfExact}) the solution
becomes singular: intensity tends to infinity asymptotically when the light propagation
distance approaches $z_{\rm sf}=b\sqrt{{\rm e}/ 2\alpha}$.

In case of the eikonal equations with arbitrary nonlinear refractive index we
constructed approximate analytical solutions. For the case of initial intensity
distribution given by equation (\ref{PS}), the approximate solution was compared with
the exact one (\ref{Solution1}), (\ref{Solution2}). It was shown that a value of the
self-focusing position provided by an approximate solution was very close to the
magnitude obtained from the exact formula (\ref{zsfExact}).

In the last section we considered a nonlinear Schr\"odinger equation in (1+2)
dimensions with arbitrary refractive index. An approximate symmetry group admitted by
both this equation and boundary conditions corresponding to collimated beam with
arbitrary initial intensity distribution was constructed. The solution was presented in
the form of algebraic equations (\ref{Sol3d}) which must be analyzed for each
particular form of the refractive index and the initial intensity distribution. As an
example, the case of two-term nonlinear refractive index Eq.~(\ref{nn2}) was examined
in details. We obtained an explicit formula for the self-focusing position
Eq.~(\ref{zsf3d}) and demonstrated that, for this form of the refractive index, the
beam collapse can also take place outside the beam axis.

\section*{Acknowledgment} V.F.K and L.L.T acknowledge financial support by LiMat
project. V.F.K also thanks DAAD for financial support. This work was also partially
supported by RFBR projects No. 09-01-00610a and 08-01-00291a.

\end{document}